\documentclass[aps,prb,reprint,showpacs,floatfix,citeautoscript,longbibliography,superscriptaddress]{revtex4-1}
\usepackage{graphicx}
\usepackage{bm,amsmath,amssymb,mathrsfs,dcolumn}
\usepackage[colorlinks=true, linkcolor=blue, citecolor=blue, urlcolor=blue, linktoc=page, bookmarks=false, pdfstartview={FitH}, pdfborder={0 0 0.0 [3 3]}]{hyperref} 
\usepackage[usenames]{xcolor} 
\hypersetup{pdfborder=0 0 0,colorlinks=true,citecolor=blue,linkcolor=blue}

\begin{document}
\title{Noncollinearity-modulated electronic properties of the monolayer CrI$_3$}
\author{Lingling Ren}
\author{Qian Liu}
\author{Pengxiang Xu}
\affiliation{Center for Advanced Quantum Studies and Department of Physics, Beijing Normal University, Beijing 100875, China}
\author{Zhicheng Zhong}
\affiliation{Key Laboratory of Magnetic Materials and Devices \& Zhejiang Province Key Laboratory of Magnetic Materials and Application Technology, Ningbo Institute of Materials Technology and Engineering, Chinese Academy of Sciences, Ningbo 315201, China}
\author{Li Yang}
\affiliation{Department of Physics and Institute of Materials Science and Engineering, Washington University in St. Louis, St. Louis, Missouri 63130, USA}
\author{Zhe Yuan}
\email{zyuan@bnu.edu.cn}
\author{Ke Xia}
\affiliation{Center for Advanced Quantum Studies and Department of Physics, Beijing Normal University, Beijing 100875, China}
\date{\today}
\begin{abstract}
Introducing noncollinear magnetization into a monolayer CrI$_3$ is proposed to be an effective approach to modulate the local electronic properties of the two-dimensional (2D) magnetic material. Using first-principles calculation, we illustrate that both the conduction and valence bands in the monolayer CrI$_3$ are lowered down by spin spiral states. The distinct electronic structure of the monolayer noncollinear CrI$_3$ can be applied in nanoscale functional devices. As a proof of concept, we show that a magnetic domain wall can form a one-dimensional conducting channel of in the 2D semiconductor via proper gating. This conducting channel is approximately 7 nm wide and has the carrier concentration of $10^{13}\sim10^{14}$~cm$^{-2}$.
\end{abstract}
\maketitle

\section{Introduction}
Two-dimensional (2D) materials, such as graphene \cite{Novoselov:sc04}, MoS$_2$ \cite{Mak:prl10}, etc., have attracted much attention in experimental and theoretical studies owing to their distinct electronic \cite{DasSarma:rmp11}, magnetic \cite{Karpan:prl07} and optical properties~\cite{Sun:nphon16} and the potential applications in designing nanoscale functional devices~\cite{Desai:sc16,Feng:nphon12}. The latter requires effective manipulation of the physical properties like the conductance, the band gap and the magnetic orientation. The recent discovery of 2D magnetic semiconductors~\cite{Gong:nat17,Huang:nat17} offers new opportunities in controlling the electronic properties by the magnetization~\cite{Klein:sc18} and vice versa~\cite{Huang:nnano18}. Topological spin waves in the 2D honeycomb lattice have been discovered using neutron scattering~\cite{Chen:arxiv18}.

Following the intriguing idea of van der Waals heterostructures~\cite{Geim:nat13,Novoselov:sc16}, the present studies of 2D magnetic materials are mostly focused on the ferromagnetic and antiferromagnetic coupling between neighboring layers. For instance in four layers of CrI$_3$, Song et al. observed a tunneling magnetoresistance, which was as large as 10$^5$\%~\cite{Song:sc18}. The ferromagnetic and antiferromagnetic coupling in bilayer CrI$_3$ can be artificially switched by a gate voltage~\cite{Huang:nnano18} or electrostatic doping~\cite{Jiang:nnano18}. The van der Waals spin valve has been proposed~\cite{Cardoso:prl18} consisting of bilayer graphene between two 2D ferromagnetic layers, where the electronic properties of the bilayer graphene strongly depend on the magnetic order of the two ferromagnetic layers. However, the magnetic coupling between 2D layers is much weaker than the in-plane coupling. Therefore tuning the in-plane magnetic order is more effective to modulate the electronic structure~\cite{Zhang:jmcc15,Jiang:nl18}.

In this paper, we use CrI$_3$ as an example of 2D magnetic materials and demonstrate the noncollinear magnetic order in the monolayer CrI$_3$ results in a novel modulation of its electronic properties. The physical origin of such modulation is revealed by first-principles calculations. The charge redistribution in noncollinear CrI$_3$ results in a noticeable increase of the work function compared with the collinear case. Thus both the conduction and valence bands are lowered down in the area of a magnetic domain wall or a Skyrmion. This energy shift can be employed, under proper gating, to generate local conducting channels, which can even be moved by a magnetic field or spin waves. The noncollinearity-modulated electronic structure of the monolayer CrI$_3$ may also be applied in electron-hole separation and creation of the identical quantum dots.

\begin{figure}[b]
\includegraphics[width=\columnwidth]{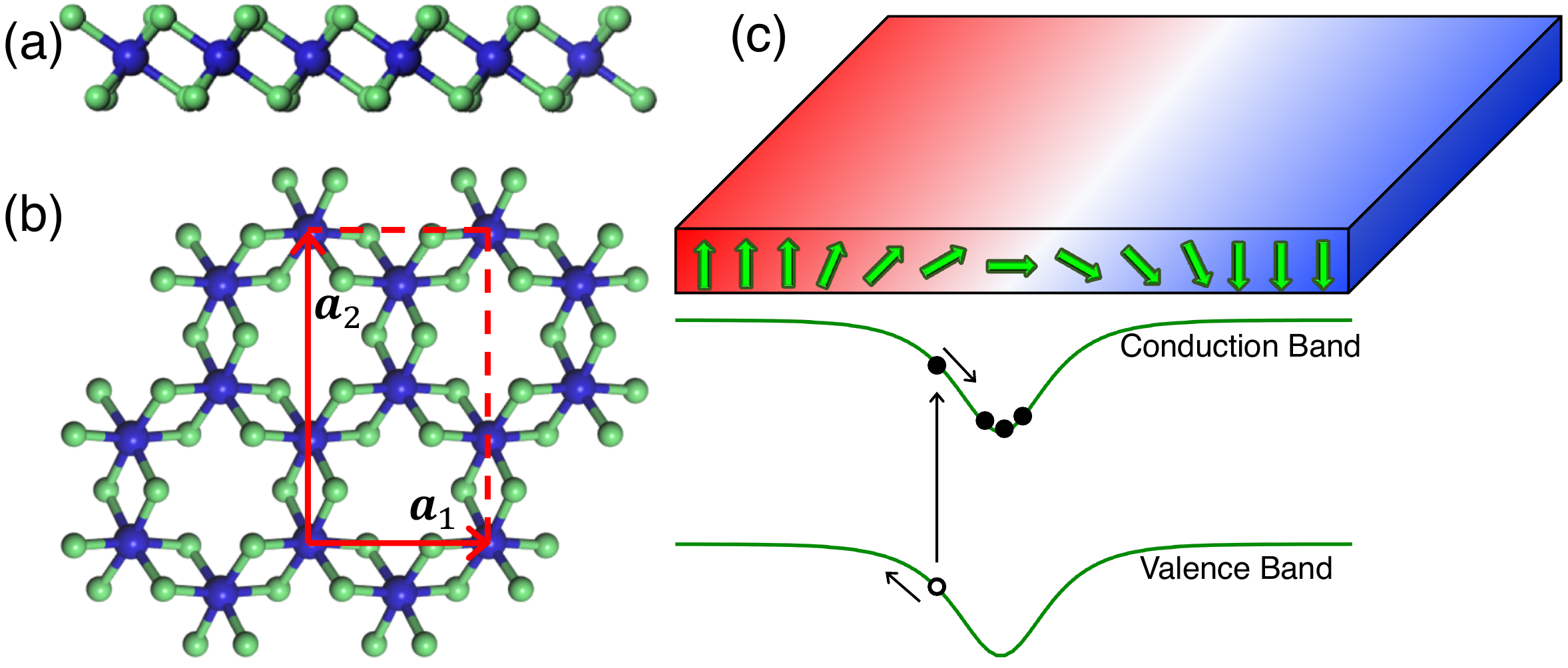}
\caption{(a) Side and (b) top views of the atomic structure of a monolayer CrI$_3$. The blue and green balls represent chromium and iodine atoms, respectively. The red arrow $\mathbf a_1$ ($\mathbf a_2$) denotes the in-plane translational vector along zigzag (armchair) direction. (c) Schematic illustration of a magnetic domain wall in a monolayer CrI$_3$ and its influence on the electronic structure. The left (red) and right (blue) magnetic domains have the upward and downward magnetization, respectively. At the domain wall (white region), the noncollinear magnetization lowers down both the conduction and valence bands resulting in potential wells along the domain wall.}\label{fig:1}
\end{figure}
\section{A prototype proposal}
The atomic structure of the monolayer CrI$_3$ is shown in Fig.~\ref{fig:1}, where chromium atoms form a 2D hexagonal lattice and iodine atoms are located at both sides of this 2D lattice. The monolayer CrI$_3$ has perpendicular magnetic anisotropy \cite{Lado:2dmater17,Webster:prb18} so the magnetization has either the upward or the downward orientation out of the 2D plane, as schematically plotted in Fig.~\ref{fig:1}(c). A magnetic domain wall is located at the boundary of the two magnetic domains, where there is a continuous transition of the magnetization orientation. As we will demonstrate later, the noncollinear magnetization in the area of a domain wall lowers down both the conduction and valence bands. Such a modulation of the local electronic structure enables novel applications of the 2D ferromagnetic material. For example, an electron can be excited into the conduction band leaving a hole in the valence band. Due to the potential gradient, the electron and hole would move towards the opposite directions leading to a natural electron-hole separation. With appropriate gating, the potential well in the conduction band that traps electrons forms a one-dimensional (1D) conducting channel in the insulating 2D sheet. In addition, since the domain wall can be moved by a magnetic field or spin waves \cite{Yan:prl11}, this conducting channel is spatially relocatable and thus offers more flexible controllability compared with other nonmagnetic 2D materials.

\section{Spin spiral states}
\begin{figure}[tb]
\includegraphics[width=\columnwidth]{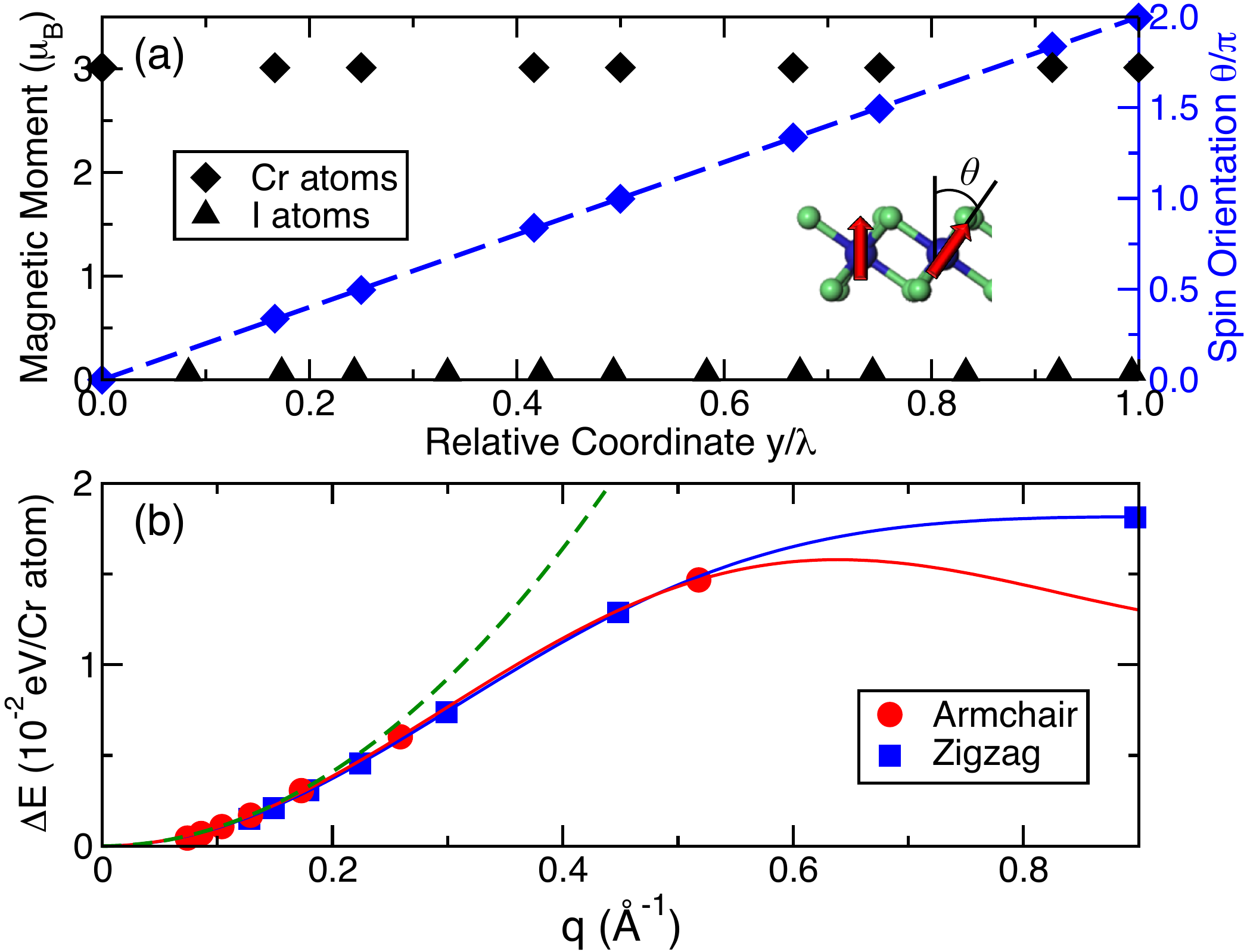}
\caption{(a) Calculated magnetic moments of Cr (black diamonds) and I atoms (black triangles) in a CrI$_3$ SS with $\lambda=24.3$~{\AA} along the armchair direction. Right axis: the calculated polar angle $\theta$ of Cr atoms (blue diamonds) as a function of the relative coordinate. The blue dashed line illustrates the linear dependence corresponding to the homogeneous magnetization gradient in the SS. Inset: sketch of the polar angle $\theta$. (b) Calculated energy difference between the SS and collinear CrI$_3$ as a function of the wave vector of the SS. The blue and red solid lines are calculated using Eq.~\eqref{eq:ediffz} and Eq.~\eqref{eq:ediffa}, respectively, with the fitted values of $J_1$ and $J_2$. The green dashed line illustrates the quadratic dependence $-(J_1/16+3J_2/8)a^2q^2$.}\label{fig:2}
\end{figure}

In real materials, the size of domain walls is determined by competition of exchange interaction and magnetic anisotropy energy (MAE). Although the domain-wall width will be estimated in Appendix~\ref{sec:A} using the calculated material parameters, we consider the magnetization gradient as an imposed parameter to study the influence of the noncollinearity on the electronic structure. Specifically, a spin spiral (SS) state is introduced in a monolayer CrI$_3$. In a SS, the magnetization varies along a certain direction with a constant spatial gradient $\nabla\theta$, where the polar angle $\theta$ is used to describe the direction of local magnetization, as illustrated in the inset of Fig.~\ref{fig:2}(a). The period of the SS is then defined by $\lambda=2\pi/\vert\nabla\theta\vert$ and its wave vector $q=2\pi/\lambda=\vert\nabla\theta\vert$. A SS essentially correspond to two 180$^{\circ}$ domain walls in contact with each other.

The calculation is carried out using the projected augmented waves method implemented in the Vienna ab initio simulation package (VASP)~\cite{Kresse:prb96,Kresse:cms96}. The general gradient approximation is employed with the Perdew-Burke-Ernzerhof (PBE) type exchange-correlation functional. The wave functions are expanded in plane wave basis with an energy cutoff of 600~eV. We sample the 2D Brillouin zone (BZ) by $10\times10$ $k$ mesh for the hexagonal primitive cell of CrI$_3$, which contains two Cr and six I atoms. By comparing to the results obtained using a much denser (80$\times$80) $k$ mesh, we have concluded that all the calculations in this work are well converged. For larger systems, we keep the same density of $k$ points unchanged in the reciprocal space. A 25-{\AA}-thick vacuum perpendicular to the monolayer is used in the calculation to avoid the interaction between neighboring unit cells. Spin-orbit coupling (SOC) is included in all the calculations unless otherwise stated. Some convergence tests are given in Appendix~\ref{sec:A}.

Figure~\ref{fig:2}(a) shows the calculated magnetic moment of every atom in CrI$_3$ with a SS along the armchair direction and $\lambda=24.3$~{\AA}, which are plotted as a function of the relative coordinate along the SS direction. Every Cr atom has a magnetic moment 3~$\mu_B$ while the magnetic moments of I atoms are negligible. These values are the same as in the collinear case. The polar angle $\theta$ of every Cr atom in the SS exhibits a linear dependence on the position. Note that both the direction and magnitude of the local magnetic moment are fully relaxed in the calculation indicating that the SS state is a metastable state. Although the SS state has a higher energy than the collinear magnetization, it can not spontaneously relax to the collinear state without an energy cost.

The energy difference between a SS state and the collinear magnetization is plotted in Fig.~\ref{fig:2}(b) as a function of the wave vector $q$. The calculated values can be well described by the exchange interaction of the nearest pairs of Cr atoms ($J_1$) and of the next nearest neighbors ($J_2$). Specifically, we can write the energy difference $\Delta E$ for zigzag and armchair SSs, respectively, as
\begin{equation}
\Delta E=\frac{J_1}{2}\left(\cos\frac{qa}{2}-1\right)+\frac{J_2}{2}\left[\cos\left(qa\right)+2\cos\frac{qa}{2}-3\right],\label{eq:ediffz}
\end{equation}
and
\begin{eqnarray}
\Delta E&=&\frac{J_1}{4}\left(\cos\frac{qa}{\sqrt{3}}+2\cos\frac{qa}{2\sqrt{3}}-3\right)\nonumber\\
&&+J_2\left(\cos\frac{\sqrt{3}qa}{2}-1\right).\label{eq:ediffa}
\end{eqnarray}
Here $a=7.0$~{\AA} is the lattice constant, which equals to the length of the vector $\mathbf a_1$ in Fig.~\ref{fig:1}. By fitting the calculated values of both zigzag and armchair SSs, we obtain the unified exchange parameters $J_1=-10.5\pm0.6$~meV and $J_2=-3.8\pm0.2$~meV, as the blue and red solid lines in Fig.~\ref{fig:2}(b). At small $q$, both Eq.~\eqref{eq:ediffz} and Eq.~\eqref{eq:ediffa} are reduced to be a unified quadratic function $\Delta E(q)=-(J_1/16+3J_2/8)a^2q^2$, as illustrated by the green dashed line in Fig.~\ref{fig:2}(b). 

It is worth noting that although SOC is included in our calculation, the MAE is much smaller than the exchange energy and is neglected in Eqs.~\eqref{eq:ediffz} and \eqref{eq:ediffa}. This approximation can be justified as follows. The total MAE of a SS with the unit cell length of one period $\lambda$ and width $W$ can be integrated over the unit cell $-K\,W \int_0^{\lambda} dx\,\cos^2\left(\frac{2\pi}{\lambda}x\right)=-\frac{KW\lambda}{2}$. Here $K=3.6$~meV/nm$^2$ is the MAE calculated in Appendix~\ref{sec:A}. The same unit cell with collinear perpendicular magnetization has the MAE $-KW\lambda$. Since the unit cell contains $\lambda W/(\sqrt{3}a^2/4)$ Cr atoms, the change in the MAE per Cr atoms due to the presence of the SS reads $\frac{KW\lambda}{2}\cdot\frac{\lambda W}{\sqrt{3}a^2/4}=\sqrt{3}a^2 K/8\approx0.4$~meV, which is independent of the type (zigzag or armchair) and the wave vector. Therefore the magnetic anisotropy contributes a constant of approximately 0.4~meV per Cr atom into $\Delta E$, which is much smaller than the numbers ($\sim10$~meV) used in the fitting and hence can be neglected in practice. 

\begin{figure}[t]
\includegraphics[width=\columnwidth]{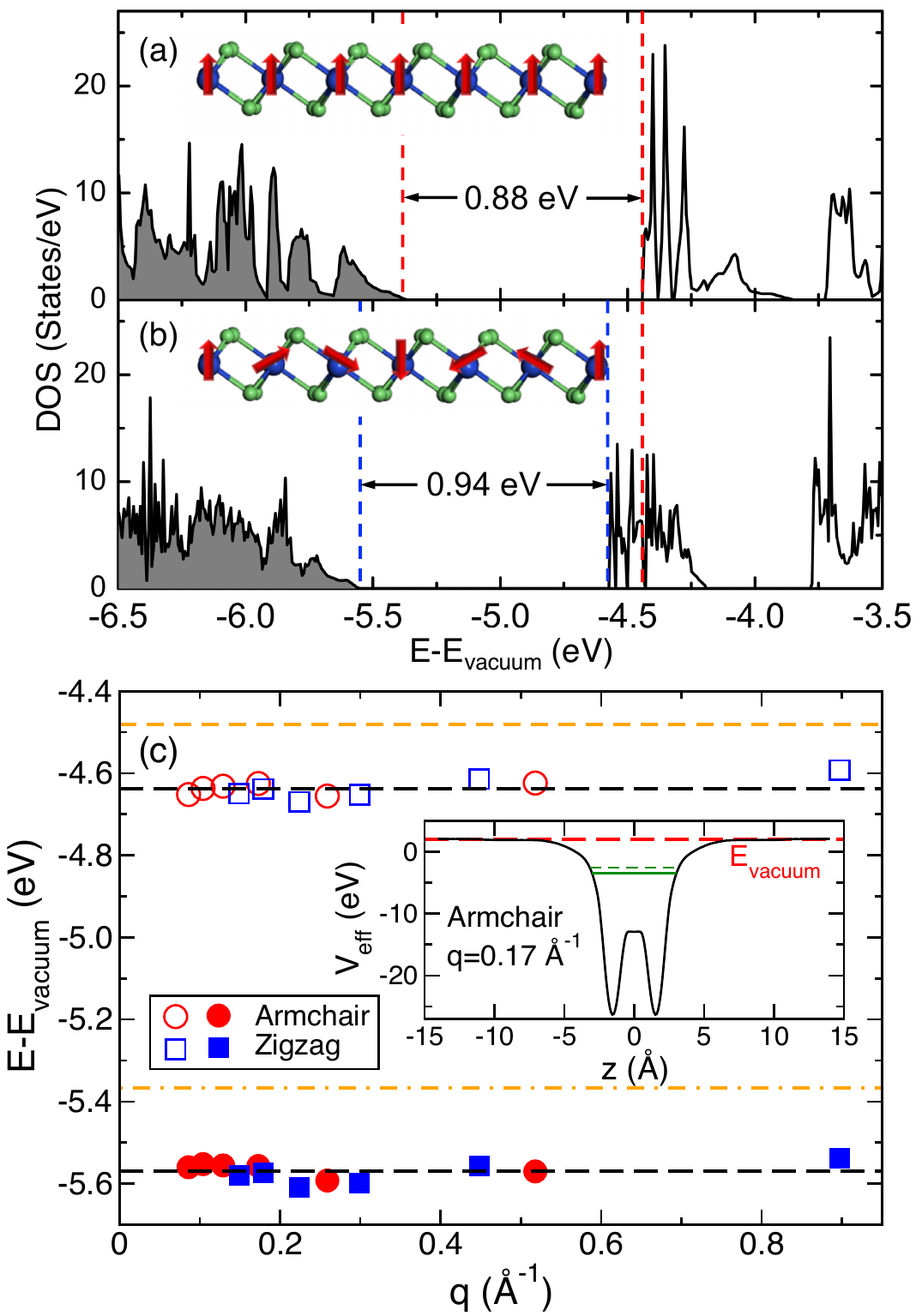}
\caption{(a) Calculated DOS of CrI$_3$ with collinear magnetization. The vacuum energy is set as the reference and the shadow at low energy represents the occupied states. The two vertical dashed lines denotes the highest energy of the occupied states and the lowest energy of the unoccupied states. (b) Calculated DOS of a zigzag SS state in CrI$_3$ with $q=0.3$~{\AA}$^{-1}$. The plotted DOS is normalized by the number of Cr atoms in the unit cell. (c) The highest energy of the occupied states (solid symbols) and the lowest energy of the unoccupied states (empty symbols) in CrI$_3$ as a function of the wave vector $q$. The black dashed lines are guides for the eyes. The orange dashed (dash-dotted) line illustrates the lowest energy of the unoccupied states (the highest energy of the occupied state) of collinear CrI$_3$. Inset: calculated effective Kohn-Sham potential for an armchair SS with $q=0.17$~{\AA}$^{-1}$ as a function of $z$ after averaging over the $xy$ plane. The red dashed line illustrates the vacuum energy and the green solid and dashed lines represent the highest energy of the occupied states and the lowest energy of the unoccupied states, respectively.}\label{fig:3}
\end{figure}
\section{Electronic properties of noncollinear CrI$_3$}
To study the influence of noncollinear magnetization on the electronic structure, we calculate the density of states (DOS) of CrI$_3$ without and with a SS state, which are shown in Fig.~\ref{fig:3}(a) and (b), respectively. For collinear case, the band gap is 0.88~eV and the work function is approximately 5.35~eV. The latter is defined by the energy difference between the highest energy of occupied states and the vacuum energy; see the inset of Fig.~\ref{fig:3}(c). For the zigzag SS state with $q=0.3$~{\AA}$^{-1}$, the band gap has a slight variation of less than 0.1~eV, but the work function has a noticeable increase of 0.2~eV. Therefore both the highest occupied state and the lowest unoccupied state are shifted towards lower energy by the noncollinear magnetization. 

The calculated energies of the highest occupied state and the lowest unoccupied state are plotted in Fig.~\ref{fig:3}(c). The energy of the lowest unoccupied state is lowered from -4.48~eV for the collinear case (the orange dashed line) down to around -4.64~eV for both armchair and zigzag SSs (the empty symbols). Moreover, this energy shift does not sensitively depend on the wave vector $q$; the empty circles and squares in Fig.~\ref{fig:3}(c) are located in a very narrow energy range of [-4.59,-4.67]~eV. Analogously, the calculated energies of the highest occupied state are independent of the type of SSs and of the wave vector $q$. Compared to the highest occupied state of collinear CrI$_3$ at -5.37~eV (the orange dash-dotted line), this energy with SSs is shifted down to approximately -5.57~eV by the presence of the noncollinear magnetization (the solid symbols). We have artificially turned off the spin-orbit interaction in the calculation and found that the above energy shift still exists. In particular, the electronic states on both sides of the band gap contains mostly the minority-spin states. Such unique properties leads to the discovered energy shift, which is noticeable but not destructive. We emphasize that the local electronic properties of a magnetic domain wall share the same features as that calculated for SSs because a SS can be divided into two 180$^\circ$ domain walls with the same magnetization gradient.

\begin{figure}[t]
\includegraphics[width=.9\columnwidth]{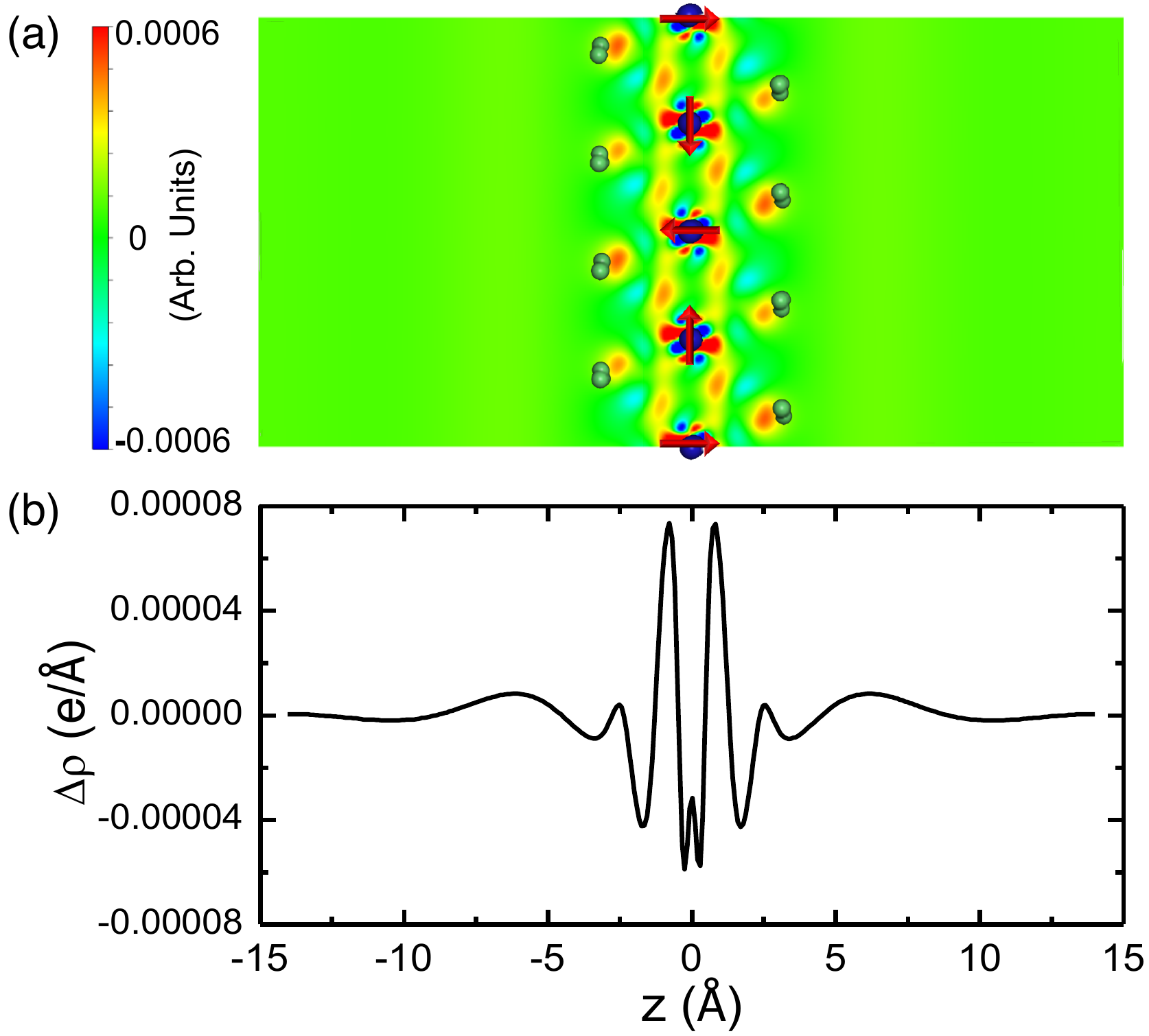}
\caption{(a) Electron density difference between noncollinear and collinear magnetization. The red (blue) color indicates the larger (lower) electron density in noncollinear CrI$_3$ compared with the collinear case. The blue and green spheres denote Cr and I atoms, respectively. The red arrows illustrate the magnetic moment orientation in the noncollinear CrI$_3$. (b) In-plane integrated electron density difference between the noncollinear and collinear CrI$_3$ as a function of the out-of-plane coordinate ($z$).}
\label{fig:4}
\end{figure}
The shift of both the conduction and valence bands of CrI$_3$ towards the lower energy caused by the noncollinear magnetization is the key result of this work. Such an unexpected energy shift originates from the noncollinearity-induced electron density redistribution, which in turn results in the increase of the work function. In Fig.~\ref{fig:4}(a), we plot the electron density difference in a cross section between the collinear CrI$_3$ and a zigzag SS state. Positive (negative) value indicates a larger (smaller) electron density in noncollinear Cr$_3$ than that of the collinear magnetization. The charge redistribution takes place mainly in the vicinity of Cr atoms and electrons move away from the central Cr plane. Near I atoms, the charge redistribution is relatively weaker and electrons also moves away from the central plane of CrI$_3$. This trend can be seen clearly if we perform the in-plane integration of the electron density difference, as shown in Fig.~\ref{fig:4}(b). The electron density redistribution essentially increases the surface dipole~\cite{Lang:prb70} on both sides of CrI$_3$. Owing to this enhanced surface dipole, the work function of noncollinear CrI$_3$ is larger than that of the collinear case. This is the physical reason why both the valence and conduction bands are shifted downwards by the noncollinear magnetization. 

\section{Discussions}
The noncollinearity-modulated electronic structure in the 2D ferromagnet CrI$_3$, which is schematically illustrated in Fig.~\ref{fig:1}(c), allows us to design nanoscale functional devices. The aforementioned mobile 1D conducting channel in an insulating 2D sheet can be realized by appropriate gating. For example, various metallic substrates may give rise to different gating effect, which has been studied in graphene and MoS$_2$ on metallic substrates~\cite{Giovannetti:prl08}. Alternatively, ionic gating can be applied and is expected to have prominent effect~\cite{Misra:apl07}. To be more realistic, we estimate the geometric width of the 1D conducting channel or equivalently the width of domain walls in the monolayer CrI$_3$. The domain-wall width is essentially determined by the spin stiffness $A$ and magnetic anisotropy energy $K$. A large $A$ leads to a slowly varying magnetization and hence a wide domain wall, whereas a large $K$ tends to align the magnetization along the easy axis as much as possible resulting in a narrow domain wall. Using the calculated values $A=-(J_1+6J_2)/\sqrt{3}=19.2\pm0.8$~meV and $K=3.6\pm1.0$~meV/nm$^{2}$ in Appendix~\ref{sec:A}, we obtain the domain-wall width in the monolayer CrI$_3$ $\lambda_{\rm DW}=\pi\sqrt{A/K}=7.2\pm1.0$~nm~\cite{Hubert:1998}. The domain walls in the monolayer CrI$_3$ are much more narrow than those in ferromagnetic 3$d$ metals and alloys and are very promising for precise manipulation of nanoscale electron transport. 

The carrier concentration in the 1D conducting channel can be estimated using the calculated DOS. As shown in Fig.~\ref{fig:3}(a) and (b), the conduction band bottom of noncollinear CrI$_3$ is lower than that in the collinear case. If the Fermi level of a monolayer CrI$_3$ is tuned between the two band bottoms, the domain wall becomes conducting while the domains with collinear magnetization are still insulating. The integration over the energy between the blue and red dashed lines in Fig.~\ref{fig:3}(b) results in 0.8627 states. Because the plotted DOS is normalized by the number of Cr atoms in the unit cell, the integral shall be divided by the area per Cr atoms, i.e. $\sqrt{3}a^2/4=0.212$~nm$^2$. Therefore, we find the electron density 
\begin{equation}
n_c=\frac{0.8627}{0.212}=4.1\,\mathrm{nm}^{-2}=4.1\times10^{14}\, \mathrm{cm}^{-2}.
\end{equation}
The electron densities $n_c$ for other wave vectors are estimated in the same manner, as listed in Table~\ref{tab1}, which show very little dependence on $q$. In practice, one can possibly charge 10\% to 50\% of the states between the blue and red dashed lines in Fig.~\ref{fig:3}(b), and the carrier concentration for the conducting channel is approximately $0.3$--$1.5\times10^{14}$~cm$^{-2}$. This carrier concentration is of the same order of magnitude as the reported experimental values for typical 2D materials. For example, the carrier density $10^{13}$~cm$^{-2}$ of monolayer and bilayer graphene has been reported \cite{Novoselov:sc04,Pallecchi:apl12,Zhu:prb09}. The carrier concentration on the order of $10^{12}$~cm$^{-2}$ in MoS$_2$ has been reported \cite{Deng:jap15} and it can be increased to be an order of magnitude larger via doping or gating~\cite{Deng:jap15,Chakraborty:prb12}. The estimated carrier concentration is 2 or 3 orders of magnitude smaller than the density of valence electrons ($2.6\times10^{16}$~cm$^{-2}$) in the monolayer CrI$_3$ such that the electronic structure can hardly be influenced by the occupation of these carrier. This fact justifies our estimation of carrier concentration based on the calculated DOS of neutral systems.
\begin{table}[t]
\caption{Calculated electron density $n_c$ in noncollinear CrI$_3$ in the conduction bands, which are in energy below the bottom of conduction band for the collinear case. In practice, we consider 10\% to 50\% of these states are occupied so the carrier concentration shall between 0.1$n_c$ and 0.5$n_c$.}
\begin{center}
\begin{tabular}{ccc}
\hline\hline
Orientation & $q$ ({\AA}$^{-1}$) & $n_c$ ($10^{14}$~cm$^{-2}$)  \\
\hline
Armchair & 0.086 & 4.2 \\
Armchair & 0.104 & 3.8  \\
Zigzag & 0.149 & 4.1 \\
Zigzag & 0179 & 3.8  \\
Zigzag & 0.224 & 4.0  \\
Zigzag & 0.299 & 4.1  \\
Zigzag & 0.448 & 3.0  \\
Armchair & 0.518 & 4.2 \\
Zigzag & 0.897 & 3.1  \\
\hline\hline
\end{tabular}
\end{center}
\label{tab1}
\end{table}

In addition, a monolayer CrI$_3$ on a substrate with strong SOC would induce the so-called Dzyaloshinskii-Moriya interaction~\cite{Dzyaloshinskii:jpcs58,Moriya:pr60} because of the broken inversion symmetry. Therefore, magnetic Skyrmions could be created or form a periodic lattice~\cite{Jiang:prep17}. These Skyrmions can be periodic potential wells of electrons because of the lowered conduction bands. With electrons trapped in these potential wells under proper gating, identical quantum dots~\cite{Li:prl02} can be artificially created in such a system. It is worth emphasizing that both the 1D conducting channel and the electron trap can be moved by applying a magnetic field or exciting spin waves. Such a mobility of the noncollinear magnetization allows us to realize the reprogrammable electronic devices at nanoscale \cite{Lan:prx15}.

\section{Conclusions}
Using first-principles calculation, we have demonstrated that the noncollinear magnetization in a monolayer of 2D ferromagnet CrI$_3$ lowers down both the conduction and valence bands and hence increases the work function of CrI$_3$. The noncollinearity-induced modulation results from the charge redistribution, which enhances the surface dipole moment of CrI$_3$. This effect suggests a new and efficient way of tuning the electronic structure of 2D ferromagnetic materials by introducing, e.g., a magnetic domain wall or a Skyrmion, and shall have potential application in designing nanoscale functional devices made of 2D materials.

\begin{acknowledgments}
This work was financially supported by National Key Research and Development Program of China (Grant No. 2017YFA0303300), the National Natural Science Foundation of China (Grants No. No. 61774017, No. 61774018, No. 11734004, and No. 21421003), the Recruitment Program of Global Youth Experts, and the Fundamental Research Funds for the Central Universities (Grant No. 2018EYT03). 
\end{acknowledgments}

\appendix
\section{Estimation of domain-wall width in the monolayer CrI$_3$}\label{sec:A}
A magnetic domain wall width is determined by the competition of two material parameters, i.e. the spin stiffness and magnetic anisotropy energy. In this appendix, we estimate the spin stiffness from the fitted exchange interaction and calculate the MAE of the monolayer CrI$_3$. 

\subsection{Spin stiffness}
By introducing a noncollinear magnetization into a magnetic system, the change in the exchange energy $\Delta E$ can be expressed in terms of the spin stiffness $A$ as \cite{Hubert:1998}
\begin{equation}
\Delta E=A\int d^3r\,\vert\nabla\mathbf m(\mathbf r)\vert^2.\label{eq:Adef}
\end{equation}
We only consider the exchange interaction of the nearest neighbors (NNs) $J_1$ and that of the next nearest neighbors (NNNs) $J_2$ of magnetic Cr atoms. Then the exchange energy change reads
\begin{equation}
\Delta E=\sum_{n=1}^N\left(\sum_{i}^{\rm NNs}J_1\mathbf m_n\cdot\mathbf m_i +\sum_{i}^{\rm NNNs}J_2\mathbf m_n\cdot\mathbf m_i-3J_1-6J_2\right),\label{eq:totalE}
\end{equation}
where $N$ is the total number of Cr atoms in the unit cell and $\mathbf m_n$ is the magnetization direction of an arbitrary Cr atom. Without loss of generality, we consider the magnetization gradient along $x$ direction. The magnetization of neighboring Cr atoms at $\mathbf r_i$ around $\mathbf m_n$ (located at $\mathbf r_n$) follows the Taylor expansion (up to the second order term)
\begin{equation}
\mathbf m_i=\mathbf m_n+\frac{d\mathbf m}{dx}(x_i-x_n)+\frac{1}{2}\frac{d^2\mathbf m}{dx^2}(x_i-x_n)^2.\label{eq:expansion}
\end{equation}
Substituting Eq.~\eqref{eq:expansion} into Eq.~\eqref{eq:totalE}, we obtain
\begin{equation}
\Delta E=N\mathbf m\cdot\frac{d^2\mathbf m}{dx^2}\frac{J_1+6J_2}{4}a^2,
\label{eq:totalE2}
\end{equation}
where we have used the following relations for the honeycomb lattice,
\begin{equation}
\sum_{i}^{\rm NNs}(x_i-x_n)=\sum_{i}^{\rm NNNs}(x_i-x_n)=0,
\end{equation}
and
\begin{equation}
\sum_{i}^{\rm NNs}(x_i-x_n)^2=\frac{a^2}{2},\,\,\,\sum_{i}^{\rm NNNs}\left(x_i-x_n\right)^2=3a^2.
\end{equation}
Because of the relation
\begin{equation}
\mathbf m\cdot\frac{d^2\mathbf m}{dx^2}+\left\vert\frac{d\mathbf m}{dx}\right\vert^2=\frac{d}{dx}\left(\mathbf m\cdot\frac{d\mathbf m}{dx}\right)=0,
\end{equation}
we can rewrite Eq.~\eqref{eq:totalE2} as
\begin{equation}
\Delta E=-\left\vert\frac{d\mathbf m}{dx}\right\vert^2 N\frac{J_1+6J_2}{4}a^2.\label{eq:totalE3}
\end{equation}
Considering the number of Cr atoms per unit area is $ \frac{4}{\sqrt{3}a^2}$, we can replace the number of Cr atoms $N$ in Eq.~\eqref{eq:totalE3} by an areal integral $N=\int d^2r\, \frac{4}{\sqrt{3}a^2}$ and finally arrive at
\begin{eqnarray}
\Delta E&=&-\int d^2r\,\left\vert\frac{d\mathbf m}{dx}\right\vert^2 \frac{4}{\sqrt{3}a^2}\left(\frac{J_1}{4}+\frac{3J_2}{2}\right)a^2\nonumber\\
&=&-\frac{J_1+6J_2}{\sqrt{3}}\int d^2r\,\left\vert\frac{d\mathbf m}{dx}\right\vert^2.
\end{eqnarray}
Comparing the above expression with the definition, Eq.~\eqref{eq:Adef}, we find the spin stiffness of the monolayer CrI$_3$ $A=-(J_1+6J_2)/\sqrt{3}=19.2\pm0.8$~meV. Note that the unit of $A$ for 2D ferromagnets is different from that of three-dimensional materials.

\begin{figure}[t]
\begin{center}
\includegraphics[width=\columnwidth]{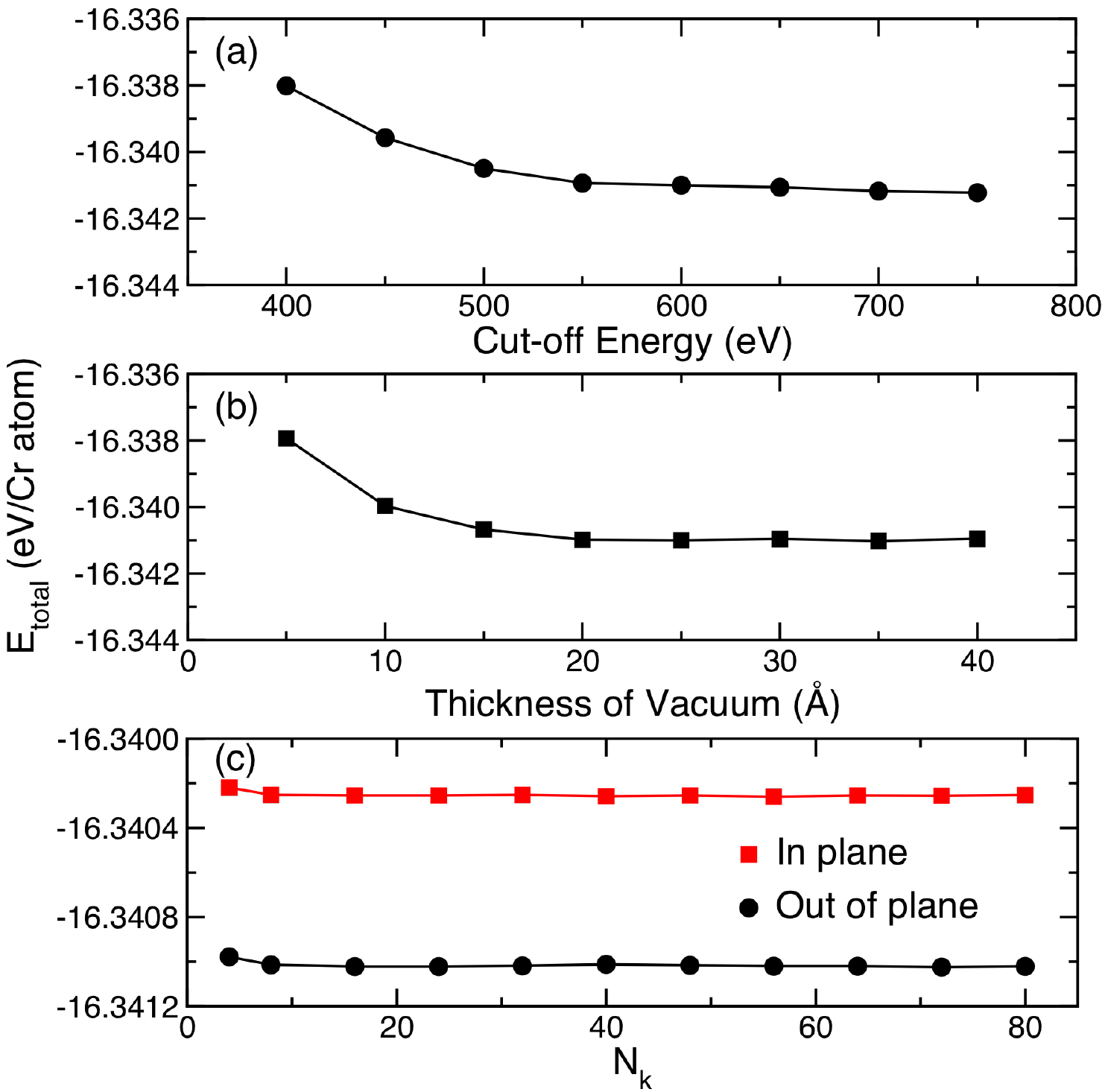}
\caption{(a) Calculated total energy as a function of cut-off energy of plane waves used in the calculation. (b) Calculated total energy as a function of thickness of vacuum perpendicular to the atomic plane, which is used to avoid artificial interactions between unit cells. Spin-orbit interaction is turned on in these numerical tests and the magnetization is collinear and perpendicular to the atomic plane. (c) Calculated total energy per Cr atom of the monolayer CrI$_3$ as a function of $N_k$, where the 2D BZ is sampled with $N_k\times N_k$ $k$-mesh. The black circles (red squares) are obtained with the magnetization perpendicular to (parallel with) the atomic plane.}
\label{fig5}
\end{center}
\end{figure}

\subsection{Magnetic anisotropy energy}
It is usually very challenging to find the converged value of the MAE using first-principles calculations and therefore we first examine some numerical convergence in our calculation. Here we consider the hexagonal unit cell of the monolayer CrI$_3$ containing two chromium atoms and six iodine atoms. Spin-orbit interaction is turned on and the collinear magnetization is set to be perpendicular to the atomic plane. The 2D BZ is sampled by a 10$\times$10 $k$ mesh. A much denser sampling with 80$\times$80 $k$ mesh has been tested later and all the reported results are well converged with respect to the sampling of $k$ points in the BZ. Figure~\ref{fig5}(a) shows the calculated total energy as a function of cut-off energy indicating the convergence is achieved at 600~eV. Since the periodic boundary condition is imposed in the calculation, we have checked the influence of the thickness of vacuum perpendicular to the atomic plane. As plotted in Fig.~\ref{fig5}(b), the total energy is converged slowly until a 20-{\AA}-thick vacuum is used. For safety, we use a 25-{\AA}-thick vacuum in our calculations.

The MAE of the monolayer CrI$_3$ is calculated by comparing the energy difference between the magnetization aligned inside the atomic plane and perpendicular to it. As shown in Fig.~\ref{fig5}(c), the calculation is carried out with very dense (up to 80$\times$80) $k$-mesh, where the magnetic anisotropy energy is extracted 0.77~meV per Cr atom, which is in good agreement with the reported values, 0.65~meV~\cite{Lado:2dmater17}, 0.80~meV~\cite{Webster:prb18}, and 0.98 meV~\cite{Jiang:nl18}. The minor difference can be attributed to the different computational methods used in the calculations. Summarizing the spread in the reported MAEs in literature into the uncertainty, we obtain the final value $0.77\pm0.20$~meV per Cr atom, or equivalently $K=3.6\pm1.0$~meV/nm$^2$.


\begin{thebibliography}{}

\bibitem{Novoselov:sc04} K. S. Novoselov, A. K. Geim, S. V. Morozov, D. Jiang, Y. Zhang, S. V. Dubonos, I. V. Grigorieva, A. A. Firsov, Electric field effect in atomically thin carbon films, Science {\bf 306}, 666 (2004).

\bibitem{Mak:prl10} Kin Fai Mak, Changgu Lee, James Hone, Jie Shan and Tony F. Heinze, Atomically thin MoS$_2$: A new direct-gap semiconductor, Phys. Rev. Lett. {\bf 105}, 136805 (2010).

\bibitem{DasSarma:rmp11} S. Das Sarma, Shaffique Adam, E. H. Hwang and Enrico Rossi, Electronic transport in two-dimensional graphene, Rev. Mod. Phys. {\bf 83}, 407 (2011).

\bibitem{Karpan:prl07} V. M. Karpan G. Giovannetti, P. A. Khomyakov, M. Talanana, A. A. Starikov, M. Zwierzycki, J. van den Brink, G. Brocks and P. J. Kelly, Graphite and Graphene as Perfect Spin Filters, Phys. Rev. Lett. {\bf 99}, 176602 (2007).

\bibitem{Sun:nphon16} Zhipei Sun, Amos Martinez and Feng Wang, Optical modulators with 2D layered materials, Nature Photonics {\bf 10}, 227 (2016).

\bibitem{Desai:sc16} S. B. Desai, S. R. Madhvapathy, A. B. Sachid, J. P. Llinas, Q. X. Wang, G. H. Ahn, G. Pitner, M. J. Kim, J. Bokor, C. M. Hu, H.-S. P. Wong, A. Javey, MoS$_2$ transistors with 1-nanometer gate lengths, Science {\bf 354}, 99 (2016).

\bibitem{Feng:nphon12} J. Feng, X. F. Qian, C.-W. Huang and J. Li, Strain-engineered artificial atom as a broad-spectrum solar energy funnel, Nature Photonics {\bf 6}, 866 (2012).

\bibitem{Gong:nat17} C. Gong, L. Li, Z. L. Li, H. W. Ji, A. Stern, Y. Xia, T. Cao, W. Bao, C. Z. Wang, Y. Wang, Z. Q. Qiu, R. J. Cava, S. G. Louie, J. Xia and X. Zhang, Discovery of intrinsic ferromagnetism in two-dimensional van der Waals crystals, Nature {\bf 546}, 265 (2017).

\bibitem{Huang:nat17} B. Huang, G. Clark, E. Navarro-Moratalla, D. R. Klein R. Cheng, K. L. Seyler, D. Zhong, E. Schmidgall, M. A. McGuire, D. H. Cobden, W. Yao, D. Xiao, P. Jarillo-Herrero and X. D. Xu, Layer-dependent ferromagnetism in a van der Waals crystal down to the monolayer limit, Nature {\bf 546}, 270 (2017).

\bibitem{Klein:sc18} D. R. Klein, D. MacNeill, J. L. Lado, D. Soriano, E. Navarro-Moratalla, K. Watanabe, T. Taniguchi, S. Manni, P. anfield, J. Fern{\'a}ndez-Rossier and P. Jarillo-Herrero, Probing magnetism in 2D van der Waals crystalline insulators via electron tunneling, Science {\bf 360}, 1218 (2018).

\bibitem{Huang:nnano18} B. Huang, G. Clark, D. R. Klein, D. MacNeill, E. Navarro-Moratalla, K. L. Seyler, N. Wilson, M. A. McGuire, D. H. Cobden, D. Xiao, W. Yao, P. Jarillo-Herrero and X. D. Xu, Nature Nanotechn. {\bf 13}, 544 (2018).

\bibitem{Chen:arxiv18} Lebing  Chen, Jae-Ho  Chung, Bin  Gao, Tong  Chen, Matthew  B. Stone, Alexander  I.  Kolesnikov, Qingzhen  Huang, and  Pengcheng  Dai, Topological spin excitations in honeycomb ferromagnet CrI$_3$, arXiv:1807.11452v2 (2018).

\bibitem{Geim:nat13} A. K. Geim and I. V. Grigorieva, Van der Waals heterostructures, Nature {\bf 499}, 419 (2013).

\bibitem{Novoselov:sc16} K. S. Novoselov, A. Mishchenko, A. Carvalho, A. H. Castro Neto, Science {\bf 353}, 461 (2016).

\bibitem{Song:sc18} T. C. Song, X. H. Cai, M. W.-Y. Tu, X. O. Zhang, B. Huang, N. P. Wilson, K. L. Seyler, L. Zhu, T. Taniguchi, K. Watanabe, M. A. McGuire, D. H. Cobden, D. Xiao, W. Yao, X. D. Xu, Giant tunneling magnetoresistance in spin-filter van der Waals heterostructures, Science {\bf 360}, 1214 (2018).

\bibitem{Jiang:nnano18} S. W. Jiang, L. Z. Li, Z. F. Wang, K. F. Mak and J. Shan, Controlling magnetism in 2D CrI$_3$ by electrostatic doping, Nature Nanotechn. {\bf 13}, 549 (2018).

\bibitem{Cardoso:prl18} C. Cardoso, D. Soriano, N. A. Garc{\'i}a-Marti{\'i}nez and J. Fern{\'a}ndez-Rossier, Van der Waals spin valves, Phys. Rev. Lett. {\bf 121}, 067701 (2018).

\bibitem{Zhang:jmcc15} Wei-Bing Zhang, Qiang Qu, Peng Zhu and Chi-Hang Lam, Robust intrinsic ferromagnetism and half semiconductivity in stable two dimensional single-layer chromium trihalides, J. Mater. Chem. C {\bf 3}, 12457 (2015).

\bibitem{Jiang:nl18} P. H. Jiang, L. Li, Z. L. Liao, Y. X. Zhao and Z. C. Zhong, Spin direction-controlled electronic band structure in two-dimensional ferromagnetic CrI$_3$, Nano Lett. {\bf 18}, 3844 (2018).

\bibitem{Lado:2dmater17}  J. L. Lado and J. Fern{\'a}ndez-Rossier, On the origin of magnetic anisotropy in two dimensional CrI$_3$, 2D Mater. {\bf 4}, 035002 (2017).

\bibitem{Webster:prb18} Lucas Webster and Jia-An Yan, Strain-tunable magnetic anisotropy in monolayer CrCl$_3$, CrBr$_3$, and CrI$_3$, Phys. Rev. B {\bf 98}, 144411 (2018).

\bibitem{Yan:prl11} P. Yan, X. S. Wang, and X. R. Wang, All-magnonic spin-transfer torque and domain wall propagation, Phys. Rev. Lett. {\bf 107}, 177207 (2011).

\bibitem{Kresse:prb96} G. Kresse and J. Furthm{\"u}ller, Efficient iterative schemes for ab initio total-energy calculations using a plane-wave basis set, Phys. Rev. B {\bf 54}, 11169 (1996).

\bibitem{Kresse:cms96} G. Kresse and J. Furthm{\"u}ller, Efficiency of ab-initio total energy calculations for metals and semiconductors using a plane-wave basis set, Comput. Mater. Sci. {\bf 6}, 15 (1996).

\bibitem{Lang:prb70} N. D. Lang and W. Kohn, Theory of metal surfaces: charge density and surface energy, Phys. Rev. B {\bf 1}, 4555 (1970); Theory of metal surfaces: Work function, \emph{ibid} {\bf 3}, 1215 (1971).

\bibitem{Giovannetti:prl08} G. Giovannetti, P. A. Khomyakov, G. Brocks, V. M. Karpan, J. van den Brink, and P. J. Kelly, Doping Graphene with Metal Contacts, Phys. Rev. Lett. {\bf 101}, 026803 (2008).

\bibitem{Misra:apl07} Rajiv Misra, Mitchell McCarthy, and Arthur F. Hebard, Electric field gating with ionic liquids, Appl. Phys. Lett. {\bf 90}, 052905 (2007).

\bibitem{Hubert:1998} A. Hubert and R. Sch{\"a}fer, {\it Magnetic Domains} (Springer-Verlag, Berlin Heidelberg, 1998). 

\bibitem{Pallecchi:apl12} E. Pallecchi, M. Ridene, D. Kazazis, C. Mathieu, F. Schopfer, W. Poirier, D. Mailly, and A. Ouerghi, Observation of the quantum Hall effect in epitaxial graphene on SiC(0001) with oxygen adsorption, Appl. Phys. Lett. {\bf 100}, 253109 (2012).

\bibitem{Zhu:prb09} Wenjuan Zhu, Vasili Perebeinos, Marcus Freitag, and Phaedon Avouris, Carrier scattering, mobilities, and electrostatic potential in monolayer, bilayer, and trilayer graphene, Phys. Rev. B {\bf 80}, 235402 (2009).

\bibitem{Deng:jap15} Xue-Yong Deng, Xin-Hua Deng, Fu-Hai Su, Nian-Hua Liu, and Jiang-Tao Liu, Broadband ultra-high transmission of terahertz radiation through monolayer MoS$_2$, J. Appl. Phys. {\bf 118}, 224304 (2015).

\bibitem{Chakraborty:prb12} Biswanath Chakraborty, Achintya Bera, D. V. S. Muthu, Somnath Bhowmick, U. V. Waghmare, and A. K. Sood, Symmetry-dependent phonon renormalization in monolayer MoS$_2$ transistor, Phys. Rev. B {\bf 85}, 161403(R) (2012).

\bibitem{Dzyaloshinskii:jpcs58} I. E. Dzyaloshinskii, A thermodynamic theory of ``weak'' ferromagnetism of antiferromagnetics, J. Phys. Chem. Solids, {\bf 4}, 241 (1958).

\bibitem{Moriya:pr60} T.  Moriya, Anisotropic  Superexchange  Interaction  and  Weak  Ferromagnetism, Phys. Rev. {\bf 120}, 91 (1960).

\bibitem{Jiang:prep17} Wanjun Jiang, Gong Chen, Kai Liu, Jiadong Zang, Suzanne G. E. te Velthuise, Axel Hoffmann, Skyrmions in magnetic multilayers, Phys. Rep. {\bf 704}, 1 (2017).

\bibitem{Li:prl02} J.-L. Li, J.-F. Jia, X.-J. Liang, X. Liu, J.-Z. Wang, Q.-K. Xue, Z.-Q. Li, J. S. Tse, Z. Y. Zhang and S. B. Zhang, Spontaneous assembly of perfectly ordered identical-size nanocluster arrays, Phys. Rev. Lett. {\bf 88}, 066101 (2002).

\bibitem{Lan:prx15} Jin Lan, Weichao Yu, Ruqian Wu, and Jiang Xiao, Spin-Wave Diode, Phys. Rev. X {\bf 5}, 041049 (2015).

\end{thebibliography}
\end{document}